\documentclass[12pt,preprint]{aastex}

\usepackage{graphicx}
\usepackage{dcolumn}
\usepackage{amsmath}

\usepackage{natbib}

\input{colordvi.tex}

\def\e{\epsilon}
\def\g{\gamma}
\begin{document}
\title{Probing the Nature of the Weakest Intergalactic Magnetic Fields
\\with the High-Energy Emission of Gamma-Ray Bursts}
\author{Kiyotomo Ichiki\altaffilmark{1}}
\affil{%
$^1$Research Center for the Early Universe, University of Tokyo, 7-3-1
Hongo, Bunkyo-ku, Tokyo 113-0033, Japan
}
\email{ichiki@resceu.s.u-tokyo.ac.jp}

\author{Susumu Inoue\altaffilmark{2}}
\affil{%
$^2$National Astronomical Observatory of Japan, Mitaka, Tokyo 181-8588, Japan
}
\email{inoue@th.nao.ac.jp}

\and

\author{Keitaro Takahashi\altaffilmark{3}}
\affil{%
$^3$Yukawa Institute for Theoretical Physics, Kyoto 606-8502, Japan
}
\email{keitaro@yukawa.kyoto-u.ac.jp}
\date{\today}
\begin{abstract}
 We investigate the delayed, secondary GeV-TeV emission of gamma-ray bursts
 and its potential to probe the nature of intergalactic magnetic fields.
 Geometrical effects are properly taken into account for the time delay between
 primary high-energy photons and secondary inverse Compton photons from
 electron-positron pairs, which are produced in $\gamma$-$\gamma$
 interactions with 
 background radiation fields and deflected by intervening magnetic fields. 
 The time-dependent spectra of the delayed emission are evaluated
 for a wide range of magnetic field strengths and redshifts.
 The typical flux and delay time of secondary photons from bursts at $z
 \sim 1$ are respectively $\sim 10^{-8}$ GeV cm$^{-2}$ s$^{-1}$ and $\sim 10^4$ s 
 if the field strengths are $\sim 10^{-18}$ G, as might be the case
 in intergalactic void regions. 
 We find crucial differences between the cases of coherent and tangled magnetic fields,
 as well as dependences on the field coherence length.
\end{abstract}


\maketitle


\section{Introduction}
Magnetic fields are ubiquitous in the universe, being observed not
only in stars but also in larger systems, such as spiral galaxies
\citep{1996Natur.379...47B},  
elliptical galaxies \citep{1996MNRAS.279..229M}, and clusters of galaxies
\citep{1990ApJ...355...29K,2002ARA&A..40..319C}
\citep[for a review, see][]{2002RvMP...74..775W}. 
It has been claimed recently that even superclusters may have associated
magnetic fields \citep{2006ApJ...637...19X}. So far, many observations
demonstrate that magnetic fields are common in the structured regions.

On even larger scales, magnetic fields are yet to be observationally determined. 
A well-known technique to probe large-scale magnetic fields
utilizes Faraday rotation, i.e. rotation of the polarization vector of
radiation from background sources by intervening magnetized plasma.  
For example, an upper limit of $6\times 10^{-11}$G has been derived for
a uniform magnetic field component in the intergalactic medium 
\citep{1990ApJ...360....1V}\citep[see also][]{1994RPPh...57..325K}. If intergalactic
magnetic fields originate from epochs earlier than cosmic recombination,
anisotropies in the cosmic microwave background (CMB) can also provide
constraints of nanogauss levels 
\citep{1997PhRvL..78.3610B,2006ApJ...646..719Y}.

Various theoretical possibilities exist for the origin of magnetic
fields on very large scales.  
During the inflationary epoch, magnetic fields on cosmological scales can be generated
if conformal invariance of electromagnetic interactions is violated
\citep{1988PhRvD..37.2743T,1992ApJ...391L...1R,2007JCAP...02...30B,2007PhRvD..75h3516B}.
Large-scale magnetic fields can also arise through the evolution of
primordial density fluctuations before cosmic recombination
\citep{2005PhRvL..95l1301T,2006Sci...311..827I, 
2007astro.ph..1329I,2007PhRvD..75j3501K,2007arXiv0710.4620T,2000astro.ph..5380H,
2004APh....21...59B,2005PhRvD..71d3502M}.
Although the field amplitudes are not expected to be large, $\lesssim 10^{-20}$ G,
their generation is inevitable, being induced by the observed density fluctuations.
Even after recombination, it has been proposed that magnetic fields of $\sim 10^{-20}$-$10^{-16}$ G
can emerge at cosmic reionization fronts around redshift $z\sim 15$,
through either the Biermann battery mechanism \citep{2000ApJ...539..505G}
or radiation drag effects \citep{2005A&A...443..367L}.
Such large-scale magnetic fields
originating at high $z$ may remain very weak in some intergalactic
regions at lower $z$, well below the upper limits from Faraday rotation,
particularly in void regions where the activity of astrophysical objects
is minimal \citep[e.g.][]{2006MNRAS.370..319B}.

\cite{1995Natur.374..430P} proposed a promising method to observationally probe
such weak intergalactic magnetic fields, relying on the delayed, secondary high-energy
emission from gamma-ray bursts (GRBs).
If primary photons of the GRB prompt emission have sufficiently high energies,
they can interact with photons of the cosmic infrared background (CIB) or CMB 
to create electron-positron pairs far away from the GRB
\cite[e.g.][]{1992ApJ...390L..49S,2004A&A...413..807K}.
These charged particles are then deflected by intergalactic magnetic fields
before emitting secondary GeV-TeV gamma-rays via inverse Compton (IC) upscattering of the CMB,
which reach the observer with a time delay relative to the primary gamma-rays.
This delayed emission depends on the properties of the intervening magnetic fields
and hence constitute a valuable probe of its nature. 
\citet[][hereafter RMZ04]{2004ApJ...613.1072R}
have discussed the time-dependent spectra of such delayed emission components
based on simple estimates of the delay timescale
\citep[see also][for similar and related discussions]
{2002ApJ...580.1013D,2004ApJ...604..306W,2006IAUJD...7E..35I,2007astro.ph..3759M,2007ApJ...656..306C}.
Although observational information on GeV-TeV emission from GRBs has been limited so far
\citep[e.g.][]{2001AIPC..558..383D,2007ApJ...667..358A},
significant advances are expected soon from the new generation of satellite-
and ground-based facilities that have recently begun operation
or will come online in the near future.

Here we reconsider this problem,
with the aim of achieving a more physically consistent picture of the delayed emission, as well as
clarifying the dependence on the magnetic field properties in greater detail.
First, we develop a new formalism that properly accounts for geometrical effects in the delayed radiation,
with important consequences for the time-dependent spectra.
Second, whereas previous studies treated only coherent magnetic fields,
we discuss cases of both coherent and tangled magnetic fields, and the resulting crucial differences.
We also explore the dependence on redshift through the evolution of the CIB, 
adopting the CIB models of \cite{2002A&A...386....1K,2004A&A...413..807K}.
The observational prospects are briefly addressed.

In \S 2, we describe our new formulation.
The resultant spectra of the delayed emission are presented in \S 3.
We provide a discussion in \S 4, and conclude with a summary in \S 5.

\section{Formulation}

\subsection{Preliminaries\label{section:preliminaries}}

Our new formalism for the delayed gamma-ray emission from GRBs
employs a simple description of the GRB prompt emission
\cite[see, e.g.][for reviews on GRBs]{2005RvMP...76.1143P,2006RPPh...69.2259M},
together with an adaptation of a treatment for delayed radiation due to scattering effects
in the ambient medium \citep[e.g.][]{2003A&A...399..505S}.
Following RMZ04,
we assume a simple power-law spectrum
of photon index $\alpha$ for the high-energy part of the prompt emission,
\begin{equation}
\frac{d^2 N_\gamma}{dE_\gamma dt}
= \frac{(\alpha-1)L_{\gamma,{\rm iso}}(t)}{4 \pi D_L^2 E^2_{\gamma, {\rm pk}}}
  \left(\frac{E_{\gamma}}{E_{\gamma,{\rm pk}}}\right)^{-\alpha},
  \hspace{10mm} (E_{\gamma,{\rm pk}} < E_\gamma < E_{\rm cut})~,
\label{eq:Band_Spectrum}
\end{equation}
where $L_{\gamma,{\rm iso}}(t)$ is the isotropic-equivalent gamma-ray
luminosity in the observer frame,
$E_{\gamma,{\rm pk}}$ is the observed spectral peak energy,
typically a few hundred keV,
and $E_{\rm cut}$ is the maximum cutoff energy.
We allow for time-dependence in the luminosity,
although the spectral shape is assumed to be constant for simplicity.
The luminosity distance is given by
\begin{equation}
D_L(z)=\frac{(1+z)}{H_0}\int_0^z dz' \left[\Omega_r(1+z')^4 +\Omega_m(1+z')^3
 + \Omega_\Lambda \right]^{-1/2}
\end{equation}
where $\Omega_r$, $\Omega_m$ and $\Omega_\Lambda$ are the densities of
relativistic particles, non-relativistic matter and cosmological constant
in units of the critical density, respectively, and $H_0$ is the Hubble constant.

The primary high energy photons $\gtrsim$100 GeV interact with CIB and CMB photons
to create high energy electron-positron pairs ($e^\pm$). These $e^\pm$ are responsible
for the delayed emission of gamma-rays through IC upscattering of CMB photons.
By assuming that each electron or positron (hereafter simply ``electron'')
share half the energy of the incident primary photon, $m_e \gamma_e = E_\gamma/2$,
the 
flux of $e^\pm$ corresponding to time $t=t_{\rm GRB}$ during the burst
duration can be written as
\begin{equation}
\frac{dN_{e,{\rm 0}}}{d\gamma_e dt_{\rm GRB}}
= \frac{L_{\gamma,{\rm iso}}(t_{\rm GRB})}{2\pi D_L^2}
  \frac{\alpha-1}{(2m_e)^{\alpha-1}}
  \frac{\gamma_e^{-\alpha}}{E_{\gamma,{\rm pk}}^{2-\alpha}}
  \left[1-\exp(-\tau(2\gamma_e m_e))\right]~,
\label{eq:dN_isodge}
\end{equation}
where $\tau(E_\gamma)$ is the optical depth to pair production with the CIB and CMB
for gamma-rays of energy $E_\gamma$.

\begin{figure}[ht]
\vspace*{0.2cm}
 \centering
  \rotatebox{0}{\includegraphics[width=0.7\textwidth]{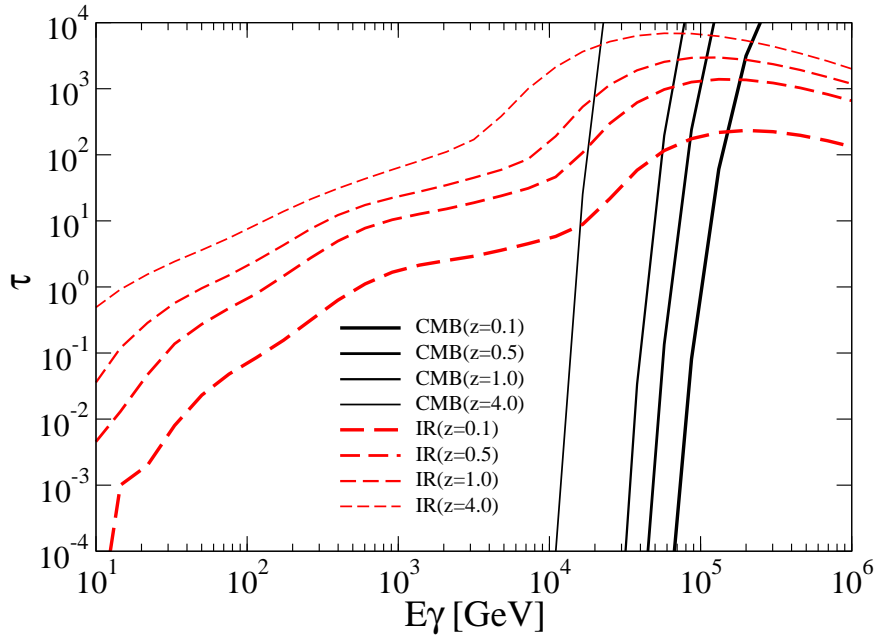}}
\caption{Optical depths due to the cosmic infrared background (IR;
 red-dashed) and cosmic microwave background (CMB; black-solid) at
 redshifts $z=0.1$ to $4.0$. } 
\label{fig:opt_depth}
\end{figure}

While the number density of CMB photons is known precisely, that of the CIB arising from
stellar and dust-reprocessed emission is still uncertain. 
\cite{2002A&A...386....1K,2004A&A...413..807K} have
developed semi-empirical, backward  evolution 
models of the CIB based on observed data from galaxy surveys.
We adopt their ``high-stellar-UV'' model which gives the best  
description
of the observed proximity effect in high-$z$ quasar absorption line  
systems,
even though their ``best-fit'' model also leads to essentially the  
same end results.
In Fig. \ref{fig:opt_depth}, we show the optical
depth to pair production employed in this paper. Note that the two peaks in the CIB curves
each correspond to the contributions from stars and dust.

Once one obtains the $e^\pm$ energy distribution,
the spectrum of the delayed emission can be calculated as
\begin{equation}
\frac{d^2 N_{\rm delayed}}{dt dE_\gamma}=\int d\gamma_e
 \frac{dN_e}{d{\gamma_e}}\frac{d^2 N_{\rm IC}}{dt dE_\gamma}~.
\label{eq:delayed_emission}
\end{equation}
Here $dN_e/d{\gamma_e}$ is the total time-integrated flux of $e^\pm$ responsible
for the delayed emission observed at time $t_{\rm obs}$ after the burst, and 
\begin{equation}
\frac{d^2 N_{\rm IC}}{dt dE_\gamma}=\frac{3\sigma_T}{4\gamma_e^2}c
\int d\e_{\gamma,{\rm CMB}} n_{\rm
CMB}(\e_{\gamma,{\rm CMB}})\frac{f(x)}{\e_{\gamma,{\rm CMB}}}~,
\label{eq:compton_power}
\end{equation}
represents the IC power from a single electron in the Thomson regime,
where $\e_{\rm CMB}$ and $n(\e_{\rm CMB})$ are the energy and number density
per unit energy of CMB photons, respectively. The function $f(x)$ is defined as
\citep{1970RvMP...42..237B}
\begin{eqnarray}
f(x)&=&2x\ln x + x+1-2x^2~,\\
 x&=&\frac{E_\gamma}{4\gamma_e^2 \e_{\rm CMB}}~.
\end{eqnarray}

At this point we need to relate $dN_{e,{\rm 0}}/d{\gamma_e}$ in Eq.(\ref{eq:dN_isodge})
with $dN_e/d{\gamma_e}$ in Eq.(\ref{eq:delayed_emission}).
RMZ04 use a simple approximation and assume
$dN_e/d{\gamma_e}=(t_{\rm IC}/\Delta t_{\rm B}) dN_{e,{\rm 0}}/d\gamma_e$,
implying that the observer receives continuously for a typical delay time $\Delta t_{\rm B}$
all photons that were emitted by $e^\pm$ during the IC cooling time $t_{\rm IC}$.
However, a more consistent treatment should consider the proper geometrical configuration of the delayed emission.
The primary photons travel distances of several Mpc or more depending on their energy
before creating $e^\pm$, which then emit secondary photons while propagating outwards,
subject to deflection by intervening magnetic fields.
In the following we formulate a relation between $dN_{e,{\rm 0}}/d\gamma_e$ and 
$dN_e/d{\gamma_e}$, taking into account these crucial geometrical effects. 

Note that the delayed secondary emission of Eq.(\ref{eq:delayed_emission}) itself
suffers $\gamma$-$\gamma$ absorption with the CIB and CMB according to $\tau(E_\gamma)$
before reaching the observer, which we take into account.
However, we neglect additional cascading effects that involve
further generations of $e^\pm$ production and IC emission.
The validity of this approximation is addressed later in \S \ref{section:discussion}.

\subsection{Geometry of delayed emission}

\begin{figure}[ht]
\vspace*{0.2cm}
\begin{minipage}[m]{0.48\linewidth}
  \rotatebox{-90}{\includegraphics[width=0.55\textwidth]{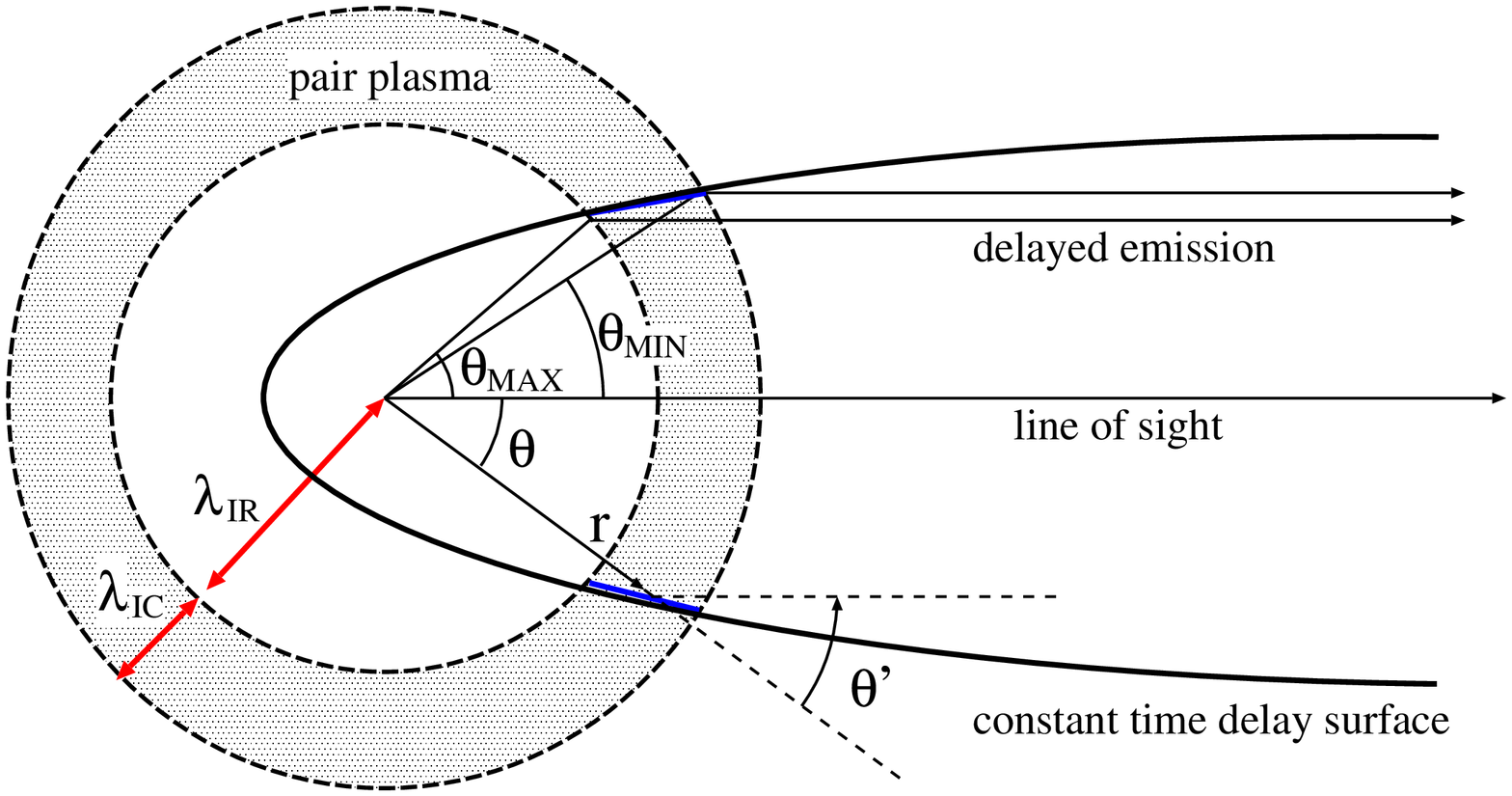}}
\end{minipage}
\begin{minipage}[m]{0.48\linewidth}
  \rotatebox{0}{\includegraphics[width=0.95\textwidth]{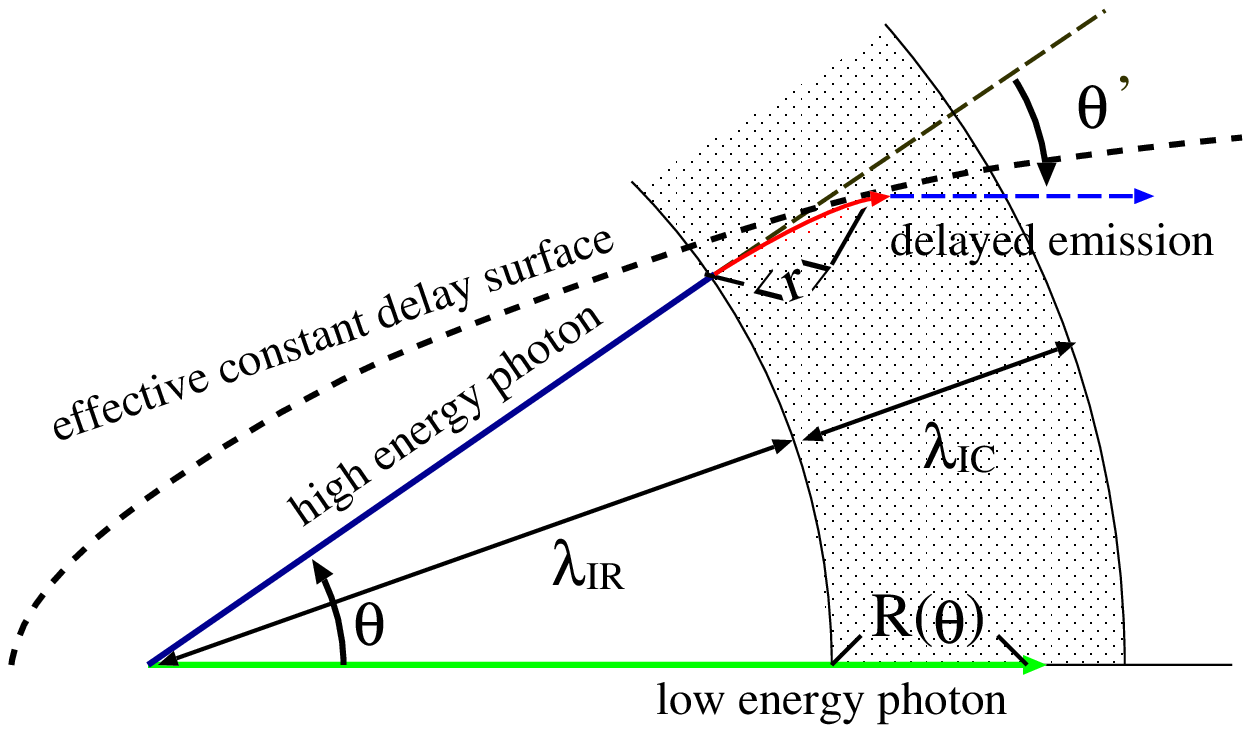}}
\end{minipage}
\caption{Schematic picture of the geometrical configuration.
 Left: Overall geometry. The angles $\theta_{\rm min}$ and $\theta_{\rm max}$ 
 delineate the range of integration in Eq.(\ref{eq:dNdgamma}).
 Right: Expanded view of the left panel. ``Effective constant delay surface''
 is the surface from which the delayed photons reach the observer at the same time,
 including the delay during propagation through intervening magnetic fields.
 See main text for details.} 
\label{fig:ITImodel}
\end{figure}

Let us suppose a GRB at redshift $z$ radiates the
spectrum of Eq.(\ref{eq:Band_Spectrum}). The geometrical configuration
is depicted in Fig. \ref{fig:ITImodel}. 
A high-energy primary photon emitted in the direction of $\theta$ with respect to the line
of sight travels a distance of $\lambda_{\rm IR}$
before interacting with a CIB photon to create an $e^\pm$.
The $\gamma$-$\gamma$ mean free path in the CIB is
\begin{equation}
\lambda_{\rm IR}
= \frac{1}{0.26 \sigma_T n_{\rm IR}}
\approx 20~{\rm Mpc} \left( \frac{n_{\rm IR}}{0.1 {\rm cm}^{-3}} \right)^{-1},
\end{equation}
where $\sigma_T$ is the Thomson cross section,
and $n_{\rm IR}$ is the number density of CIB photons
appropriate for given $E_\gamma$ and $z$,
the numerical value corresponding to $E_\gamma \approx$ 10 TeV and $z = 1$.
The electron then upscatters ambient CMB photons
with IC mean free path $\ell_{\rm IC}$ multiple times to produce secondary emission
until it loses most of its initial energy after propagating an IC cooling length $\lambda_{\rm IC}$,
which can be written
\begin{eqnarray}
&& \ell_{\rm IC}
= \frac{1}{\sigma_T n_{\rm CMB}}
\approx 1~{\rm kpc} (1+z)^{-3}, \\
&& \lambda_{\rm IC}
= \frac{3 m_e}{4 \gamma_e \sigma_T \rho_{\rm CMB}}
\approx 35~{\rm kpc} \left( \frac{E_e}{10~{\rm TeV}} \right)^{-1} (1+z)^{-4},
\label{eq:IClength}
\end{eqnarray}
where $n_{\rm CMB}$ and $\rho_{\rm CMB}$ are the number and energy density of CMB photons,
respectively.
During propagation, the electrons are deflected gradually by successive
IC scatterings as well as by the ambient magnetic fields.
The deflections can be modeled as a random walk in angle,
and the probability that an electron is deflected by an angle $\theta^\prime$ 
from its initial direction $\theta$ is
\begin{equation}
P(\theta,\theta^\prime)
= \frac{1}{\sqrt{2\pi}\sin \theta^\prime} \frac{2}{\sqrt{2\pi}\sigma(\theta)}
  \exp\left(-\frac{\left(\theta^{\prime}-\left<\theta\right>\right)^2}{2\sigma^2(\theta)}\right)~,
\end{equation}
which is normalized so that $\int P d\Omega^\prime =1$.
Here $\sigma(\theta)$ and $\left<\theta\right>$ are respectively the variance 
and expectation value of the deflection angle in the $\theta$ direction,
which encode important information about the magnetic fields
such as their amplitude and coherence length.
They are discussed below separately for the cases of tangled and coherent magnetic fields.

The secondary photons will be observed if their parent electrons have been
deflected by an angle $\theta$, and we can express $dN_e/d{\gamma_e}$ as
\begin{eqnarray}
\frac{dN_e}{d\g_e}
&=& \int dt_{\rm GRB}\int^{\theta_{\rm max}(t_{\rm obs},t_{\rm
GRB})}_{\theta_{\rm min}(t_{\rm obs},t_{\rm GRB})}
  d\theta \int d\Omega^\prime
  \delta(\theta-\theta^\prime)P(\theta^\prime,\theta)
  \frac{dN_{e,{\rm 0}}}{d\g_e dt_{\rm GRB}} \nonumber \\
&=& \int dt_{\rm GRB}\int^{\theta_{\rm max}(t_{\rm obs},t_{\rm
 GRB})}_{\theta_{\rm min}(t_{\rm obs},t_{\rm GRB})} d\theta
  \frac{2}{\sqrt{2\pi}\sigma(\theta)}
  \exp\left(-\frac{\left(\theta-\left<\theta\right>\right)^2}{2\sigma^2(\theta)}\right)
  \frac{dN_{e,0}}{d\g_e dt_{\rm GRB}}~,
\label{eq:dNdgamma}
\end{eqnarray}
where the directional requirement is represented by the delta function
$\delta(\theta-\theta^\prime)$, and the range of integration
$[\theta_{\rm min}(t_{\rm obs},t_{\rm GRB}),\theta_{\rm max}(t_{\rm
obs},t_{\rm GRB})]$
depends on the observation time (see Fig. \ref{fig:ITImodel}).
In the second equality of Eq. (\ref{eq:dNdgamma}), the integration over
$\int d\Omega^\prime = \int d\phi^\prime \sin \theta^\prime d\theta^\prime$ was performed.
The observed delayed emission can then be expressed as an integration over $dN_e/d\g_e$,
\begin{equation}
\frac{d^2 N_{\rm delayed}}{dt_{\rm obs} dE_\gamma}=\int d\g_e 
\frac{dN_e}{d\g_e}\frac{3\sigma_T}{4\g_e^2}\frac{d\left<r\right>}{dt_{\rm obs}}
\int d\e_{\gamma,{\rm CMB}} n_{\rm CMB}(\e_{\gamma,{\rm CMB}})\frac{f(x)}{\e_{\gamma,{\rm CMB}}}~.
\end{equation}
Since all photons emitted from the constant delay surface reach the observer at the same time
\citep{2003A&A...399..505S}, here we have replaced $c$ in Eq.(\ref{eq:compton_power})
with $d\left<r\right>/dt_{\rm obs}$, where $\left<r\right>$ is the
rectilinear distance between the locations of the electron at production and at $t_{\rm obs}$
(Fig. \ref{fig:ITImodel}). 

\subsubsection{Tangled magnetic fields}

If the ambient magnetic fields are tangled, that is, their coherence length is
much smaller than the electron cooling length, electrons undergo a random walk in angle
not only by IC scattering but also by magnetic deflection.
For such fields,
\begin{eqnarray}
\sigma^2(\theta)
&=& \left<\theta_B^2(\theta)\right> + \left<\theta_{\rm IC}^2(\theta)\right>+\frac{1}{\gamma_e^2}~, \label{eq:vartheta}\\
\left<\theta\right> &=& 0~. \label{eq:exptheta}
\end{eqnarray}
The last term in Eq.(\ref{eq:vartheta}) describes the effect of angular spreading
by the initial pair production interaction.
The variance of the deflection angle due to magnetic deflection $\left<\theta_B^2(\theta)\right>$
and that due to angular spreading by IC scattering $\left<\theta_{\rm IC}^2(\theta)\right>$ are given by
\begin{eqnarray}
\sqrt{\left<\theta^2_B(\theta)\right>}
&=& \frac{r_c}{r_L}\sqrt{\frac{R(\theta)}{6r_c}}~, \label{eq:def-theta_B} \\
\sqrt{\left<\theta^2_{\rm IC}(\theta)\right>}
&=& \frac{\epsilon_{\rm CMB}}{m_ec^2}\sqrt{\frac{R(\theta)}{3\ell_{\rm IC}}}~,
\label{eq:theta_B}
\end{eqnarray}
where $r_L=\gamma_e m_e c/e B$ is the Larmor radius of the electron in a magnetic field with amplitude $B$,
$r_c$ is the coherence length of the magnetic field,
and $R(\theta)$ is the distance the electron would travel in the absence of any deflection (Fig. \ref{fig:ITImodel}).
The relation between $\left<r\right>$ and $R(\theta)$ is
\begin{equation}
R - \left<r\right>
= \frac{\tau_B}{12} R \left<\phi_B^2\right> + \frac{\tau_{\rm IC}}{12} R \left<\phi_{\rm IC}^2\right>~,
\label{eq:time_delay}
\end{equation}
where $\phi_B$ and $\tau_{B}$
are respectively the deflection angle and ``optical depth'' due to magnetic deflection,
and $\phi_{\rm IC}$ and $\tau_{\rm IC}$ are those due to IC scattering
\citep[see Eq.(9) in][]{1978ApJ...222..456A}.
With the approximations
\begin{eqnarray}
&&\phi_B \approx \frac{r_c}{r_L}~, ~~\tau_B \approx \frac{R}{r_c}~,\\
&&\phi_{\rm IC} \approx \frac{1}{\gamma_e}~,~
  \tau_{\rm IC} \approx \frac{R}{\ell_{\rm IC}}~,
\end{eqnarray}
Eq.(\ref{eq:time_delay}) becomes,
\begin{eqnarray}
&& \left<r\right> = R - \frac{R}{12}\Theta^2~, \label{eq:<r>} \\
&& \Theta^2 \equiv \left<\theta^2_B(\theta)\right> + \left<\theta^2_{\rm IC}(\theta)\right>.
\end{eqnarray}
Meanwhile, $t_{\rm obs}$ is related to $R$ and $\left<r\right>$ as
\begin{equation}
\frac{t_{\rm obs}-t_{\rm GRB}}{(1+z)}=\left(R(\theta)+\lambda_{\rm
			   IR}\right)-\left(\left<r\right>+\lambda_{\rm
			   IR}\right)\cos \theta~.
\label{eq:t_obs}
\end{equation}
Using these relations and the fact that $\theta\ll 1$,
$R(\theta)$ can be expressed in terms of $t_{\rm obs}$ and $\theta$ as
\begin{eqnarray}
R&\approx& \frac{6R}{\Theta^2 \cos\theta}
\left[ - \frac{\theta^2}{2} +
\sqrt{ \left(\frac{\theta^2}{2}\right)^2
       + \frac{\Theta^2 \cos\theta}{3R}
         \left( \frac{t_{\rm obs}-t_{\rm GRB}}{1+z}-\frac{\theta^2}{2}\lambda_{\rm IR}\right)
}
\right]\\
&\approx& \frac{2}{\theta^2}
          \left(\frac{t_{\rm obs}-t_{\rm GRB}}{1+z}  - \frac{\theta^2}{2}\lambda_{\rm IR}\right).
\label{eq:R_approx}
\end{eqnarray}

For the range of integration in Eq.(\ref{eq:dNdgamma}),
the condition $\left<r\right>=R=0$ in Eq.(\ref{eq:t_obs}) together with $\theta_{\rm max}\ll 1$
gives
\begin{equation}
\theta_{\rm max} =\sqrt{\frac{2(t_{\rm obs}-t_{\rm GRB})}{\lambda_{\rm IR}(1+z)}}~.
\end{equation}
On the other hand, taking $R=\lambda_{\rm IC}$ in Eq.(\ref{eq:R_approx}) leads to
\begin{equation}
\theta_{\rm min}=\sqrt{\frac{2(t_{\rm obs}-t_{\rm GRB})}{\lambda_{\rm tot}(1+z)}}
\end{equation}
where $\lambda_{\rm tot}\equiv \lambda_{\rm IC}+\lambda_{\rm IR}$.
The expression for $d\left<r\right>/dt_{\rm obs}$ can be obtained as
follows. Differentiating Eqs. (\ref{eq:t_obs}) and (\ref{eq:<r>}) with
respect to $t_{\rm obs}$,
\begin{eqnarray}
\frac{1}{1+z}
&=& \frac{dR}{dt_{\rm obs}}-\frac{d\left<r\right>}{dt_{\rm obs}}\cos\theta ~,
    \label{eq:1/1+z}\\
\frac{dR}{dt_{\rm obs}}
&=& \left(1-\frac{1}{6} \Theta^2 \right)^{-1}\frac{d\left<r\right>}{dt_{\rm obs}}
\approx \left(1+\frac{1}{6} \Theta^2 \right)\frac{d\left<r\right>}{dt_{\rm obs}}~.
\end{eqnarray}
From these equations one obtains
\begin{equation}
\frac{d\left<r\right>}{dt_{\rm obs}}=\frac{2c}{(1+z)\left[\theta^2 + \Theta^2 / 3 \right]}~.
\label{eq:drdt}
\end{equation}
In summary, with the above formulation, the factor
\begin{equation}
{\cal S}
= \int^{\theta_{\rm max}}_{\theta_{\rm min}} d\theta 
  \frac{2}{\sqrt{2\pi}\sigma(\theta)}e^{-\frac{(\theta-\left<\theta\right>)^2}{2\sigma^2(\theta)}}
  \frac{d\left<r\right>}{dt}
\end{equation}
should be integrated, instead of $c(t_{\rm IC}/\Delta t_{\rm B})$ as in RMZ04.

In what follows, for simplicity we take in Eq. (\ref{eq:dN_isodge}) 
\begin{equation}
L_{\gamma,{\rm iso}}(t_{\rm GRB})=L_{\gamma,{\rm iso}} \delta(t_{\rm GRB}) \Delta t_{\rm GRB}~,
\label{eq:Ldelta}
\end{equation}
implying that all primary photons are emitted at $t_{\rm GRB}=0$,
and $L_{\gamma,{\rm iso}}$ is now the luminosity averaged over the 
duration of the prompt emission $\Delta t_{\rm GRB}$,
a good approximation when $t_{\rm obs}\gg \Delta t_{\rm GRB}$.


\subsubsection{Coherent magnetic fields}

If the ambient magnetic fields are coherent, or their coherence length is much larger than
the electron cooling length, they do not contribute to random walks in angle.
We assume then that the magnetic deflection can be described by a single, small angle scattering,
so that Eqs.(\ref{eq:vartheta},\ref{eq:exptheta}) are
replaced with the approximations
\begin{eqnarray}
\sigma^2(\theta) &=& \left<\theta^2_{\rm IC}(\theta)\right>+\frac{1}{\gamma_e^2}~,\\
\left<\theta\right>&=&\frac{R}{r_L}~,
\label{eq:theta_B2}
\end{eqnarray}
as well as Eq.(\ref{eq:drdt}) with
\begin{equation}
\frac{d\left<r\right>}{dt_{\rm
 obs}}=\frac{2c}{(1+z)\left[\theta^2 + \left<\theta^2_{\rm IC}(\theta)\right> / 3 \right]}~,
\end{equation}
\citep[e.g.][]{2003A&A...399..505S}.
The resulting differences in the delayed emission from the tangled field case
will be crucial for probing the nature of the magnetic fields, as discussed below.

\section{Results - spectra of delayed emission\label{section:results}}

The time-dependent spectra of the delayed secondary emission can be calculated with
the formalism developed in the previous section.
Because our main interest is in probing the state of the intervening magnetic fields,
first we fix the GRB model parameters to a fiducial set,
\begin{equation}
(L_{\gamma,{\rm iso}},\alpha,E_{\gamma,{\rm pk}}, E_{\rm cut}, \Delta t_{\rm GRB})
= (10^{53}~{\rm erg},~ 2.2,~500~{\rm keV},~10~{\rm TeV},~50~{\rm s})~,
\label{eq:GRBparameter}
\end{equation}
most of which are typically observed values for the prompt emission
\citep{1995ARA&A..33..415F,2000ApJS..126...19P}. The only exception is $E_{\rm cut}$,
for which the current observational constraints are quite poor in the GeV-TeV bands
due to the limited sensitivity and/or field of view of previous generation instruments, as well as
the strong effects of intergalactic $\gamma$-$\gamma$ absorption for typical GRB redshifts (Fig. \ref{fig:opt_depth}).
However, in the widely discussed, internal shock model of GRB prompt emission,
one can expect that at least some bursts have spectra extending to TeV energies and above
\citep[e.g.][]{2004IJMPA..19.2385Z,2004ApJ...613.1072R,2006NJPh....8..122D,2007arXiv0705.2910A},
which we assume to be the case here.
The consequences of varying the GRB parameters are discussed in \S \ref{section:discussion}.

For the cases of random magnetic fields, we set $r_c=100$ pc
as a reference value for the coherence length.
Such values may be related to the smallest scale of fields
generated during cosmic reionization by radiation drag effects \citep{2005A&A...443..367L},
or to those observed in clusters of galaxies ($\sim$ kpc; \citet{2005A&A...434...67V}).
A further, practical reason is that $r_c$ should be much smaller than the IC cooling length (Eq.\ref{eq:IClength})
in order for our random walk formulation to be valid.

\begin{figure}[ht]
\vspace*{0.8cm}
 \centering
  \rotatebox{0}{\includegraphics[width=0.7\textwidth]{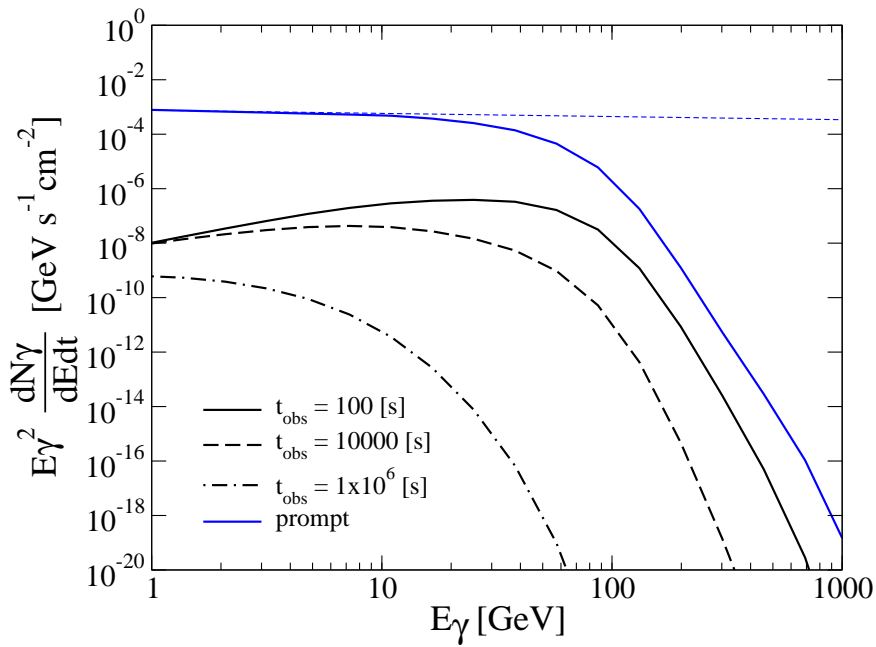}}
\caption{Spectra of delayed emission
 for a fiducial GRB at $z=1$ and $t_{\rm obs}=$100, 10$^4$, and 10$^6$ s,
 for tangled magnetic fields with $B=10^{-18}$ G and $r_c=100$ pc.
 Thin (blue) dashed and solid lines are the spectra of the prompt emission
 before and after attenuation by the CIB, respectively.} 
\label{fig:prompt}
\end{figure}

Fig. \ref{fig:prompt} shows the time-dependent spectra of the delayed
emission for a fiducial GRB at $z=1$ in the case of tangled magnetic
fields with $B=10^{-18}$ G and $r_c=100$ pc. 
The spectra generally consist of a hard power-law part at low energies
with photon indices $\la 2$, 
typical of IC emission from $e^\pm$ injected at high energies,
and a steeply falling part at high energies ($\ga 100$ GeV for $z=1$)
due to $\gamma$-$\gamma$ attenuation by the CIB.
The flux at high energy decays with time much more rapidly than at low
energy, resulting from the shorter IC cooling time as well as the
shorter delay time due to magnetic deflections for the corresponding
higher energy $e^\pm$. These properties are distinctive characteristics
of the delayed secondary emission. 

\begin{figure}[ht]
\vspace*{0.8cm}
 \centering
  \rotatebox{0}{\includegraphics[width=0.7\textwidth]{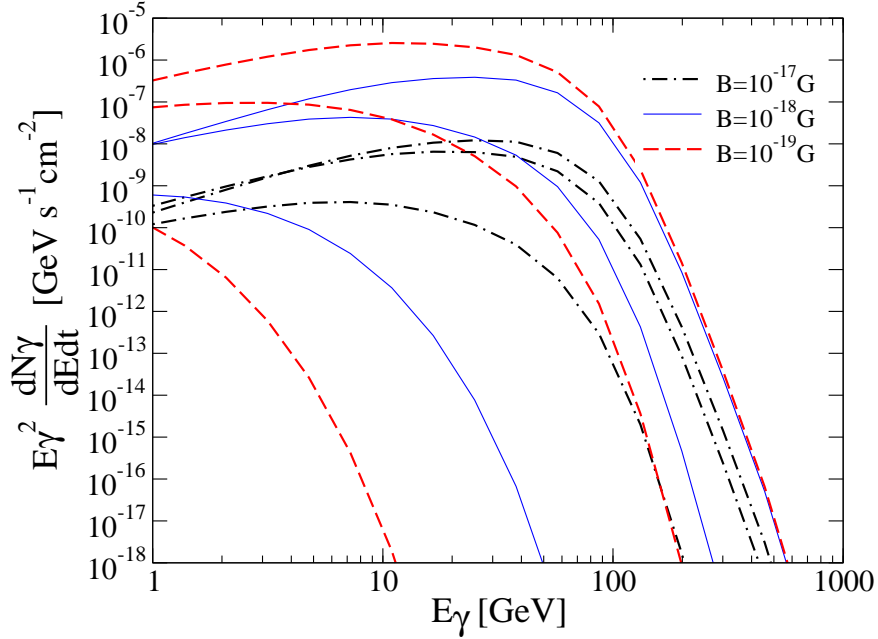}}
\caption{Spectra of delayed emission for a fiducial GRB at $z=1$
 and $t_{\rm obs}=$100, 10$^4$, and 10$^6$ s,
 for tangled magnetic fields with $r_c=100$ pc
 and $B=10^{-17}$, $10^{-18}$, and $10^{-19}$ G. 
} 
\label{fig:varyb}
\end{figure}

The delayed emission spectra for different magnetic field strengths are displayed in Fig. \ref{fig:varyb}
for the case of tangled fields with  $r_c=100$ pc.
Although changes in the intervening magnetic fields
can significantly affect the time evolution of the delayed emission,
the total fluence will remain the same.
Stronger fields induce larger delays so that the emission has lower flux at early times but lasts longer,
whereas weaker fields result in higher initial flux that decays quickly (see also RMZ04).
If the fields are sufficiently weak, $B \lesssim 10^{-20} {\rm G}$,
the delay due to angular spreading by pair creation and IC scattering dominates over magnetic deflections,
and the behavior becomes independent of $B$.
Note that $B$ and $r_c$ for tangled fields enter in our formulation
only through the combination $B \sqrt{r_c}$  (Eq. \ref{eq:def-theta_B}),
so varying $r_c$ leads to results analogous to varying $B$.

\begin{figure}[ht]
\vspace*{0.8cm}
 \centering
  \rotatebox{0}{\includegraphics[width=0.7\textwidth]{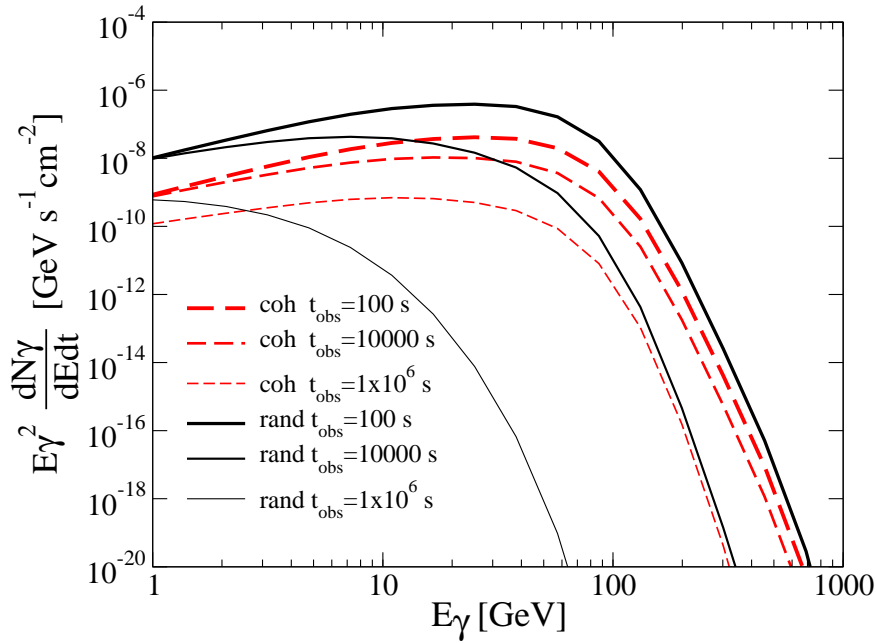}}
\caption{Spectra of delayed emission for a fiducial GRB at $z=1$ and different $t_{\rm obs}$ as indicated,
  for coherent magnetic fields with $B=10^{-18}$ G (red dashed lines),
  and tangled magnetic fields with $B=10^{-18}$ G and $r_c=100$ pc (black solid lines).
 } 
\label{fig:cohrand}
\end{figure}

Fig. \ref{fig:cohrand} is a comparison between the cases of tangled and coherent magnetic fields
with the same field strength.
We see that the spectral evolution is markedly distinct between the two cases,
owing to different dependences on $\gamma_e$ expected for the mean deflection angle:
$\sqrt{\theta_{\rm B}^2} \propto \gamma_e^{-3/2}$ for tangled fields,
as opposed to $\theta_B \propto \gamma_e^{-2}$ for coherent fields.
This offers interesting prospects for observationally probing
not only the field strength but also the coherence length,
providing valuable insight into the physical nature and origin of the intergalactic magnetic fields.

\begin{figure}[ht]
\vspace*{0.8cm}
 \centering
  \rotatebox{0}{\includegraphics[width=0.7\textwidth]{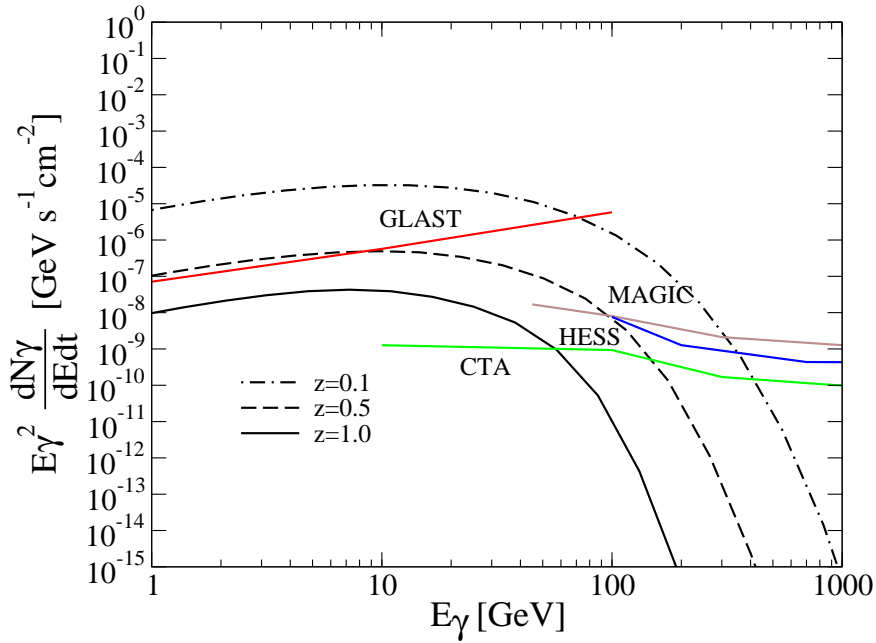}}
\caption{Spectra of delayed emission for a fiducial GRB at $t_{\rm obs}=10^4$ s
 and $z=0.1$, $0.5$, and $1.0$,
 for tangled magnetic fields with $B=10^{-18}$ G and $r_c=100$ pc.
  Overlayed are 5$\sigma$ sensitivities for GLAST, MAGIC, H.E.S.S. and CTA
  with integration times of $10^4$ s.}
\label{fig:varyz}
\end{figure}

In Fig. \ref{fig:varyz}, we depict how the spectra at fixed $t_{\rm obs}$ varies with redshift.
Besides the obvious dependence of the flux on luminosity distance,
the spectral shape of the steep, high-energy part changes considerably
due to large differences in the amount of $\gamma$-$\gamma$ attenuation.
Also shown in Fig. \ref{fig:varyz} are the sensitivies for GLAST \footnote{http://glast.gsfc.nasa.gov},
and for the Cherenkov telescopes MAGIC, H.E.S.S. and CTA \citep{2007arXiv0712.0315D}.
Note that these are actually limiting fluxes for steady sources with the given integration times,
but should serve as guides to the detectability, considering that the delayed emission
decreases monotonically with time and the fluxes at earlier times are always larger.
The observational prospects are interesting for GLAST and current Cherenkov telescopes
for bursts at $z<1$, and even at $z \ga 1$ for the future CTA project.

Finally, let us comment on the difference of our results with those of RMZ04.
They evaluate the typical delay timescale due to magnetic deflection as $\Delta t_{\rm B}=t_{\rm IC}\theta_{\rm B}^2/2$,
where $\theta_{\rm B}=\lambda_{\rm IC}/r_{\rm L}$
is the deflection angle in a coherent magnetic field.
More appropriately, however, the typical delay timescale should be given by
$\Delta t_{\rm B}=(t_{\rm IC}+\lambda_{\rm IR}/c)\theta_{\rm B}^2/2$,
because the primary high-energy photons propagate a significant distance of $\lambda_{\rm IR}$
before creating $e^\pm$ (Fig. \ref{fig:ITImodel}).
Neglecting this crucial geometrical aspect will underestimate
the effects of magnetic deflection on the delayed secondary emission.

\begin{figure}[ht]
\vspace*{0.8cm}
 \centering
  \rotatebox{0}{\includegraphics[width=0.7\textwidth]{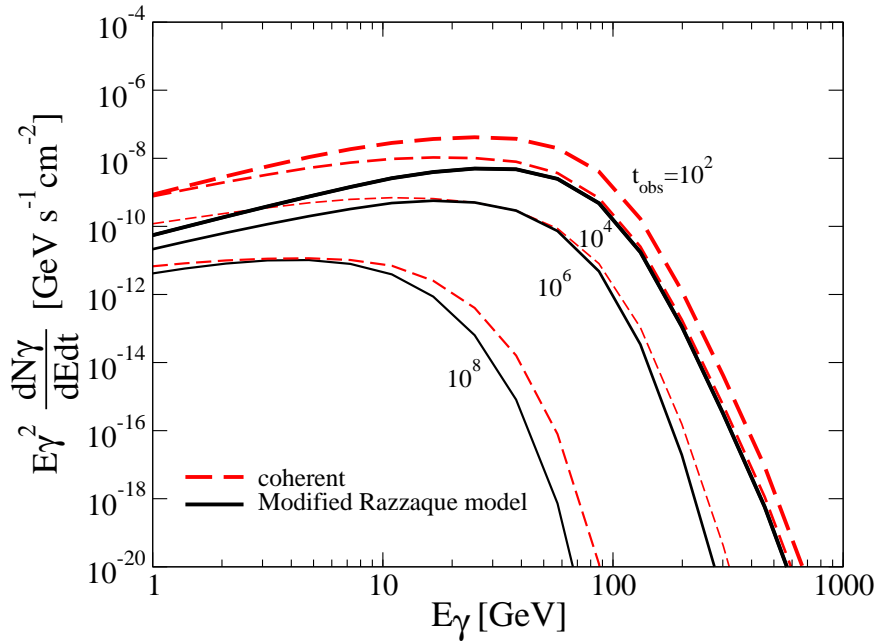}}
\caption{Spectra of delayed emission according to a modified version of RMZ04 (black solid lines)
compared with our full calculation (red dashed lines),
for a fiducial GRB at $z=1$ and $t_{\rm obs}=10^2$, $10^4$, $10^6$,
for coherent magnetic fields with $B=10^{-18}$ G.
Note that the RMZ04 curves for $t_{\rm obs} = 10^2$ and $10^4$ are indistinguishable.
} 
\label{fig:Razzaque}
\end{figure}

We have calculated the delayed emission spectra in a modified version of the RMZ04 formulation
by replacing $\Delta t_{\rm B}=t_{\rm IC} \theta_B^2$
with $\Delta t_{\rm B}=(t_{\rm IC}+\lambda_{\rm IR}/c)\theta_B^2/2$.
The redshift evolution of the CMB and CIB are also considered,
which we believe were not properly accounted for in RMZ04.
The results are compared with those of our full calculation for coherent magnetic fields in Fig. \ref{fig:Razzaque},
showing that they are broadly consistent with each other,
despite some remaining differences at low energies and at early times.

\section{Discussion\label{section:discussion}}

As demonstrated above,
the spectral evolution of the delayed secondary emission from GRBs
depends not only on the strength of the intervening magnetic fields,
but also strongly on whether the fields are tangled or coherent,
and for the former case, on the coherence length as well.
In general, magnetic fields comprise components
with a range of coherence lengths, characterized by its power spectrum.
When the spectrum can be approximated by a simple power-law,
the type of field component that determines the magnetic deflection angle
depends on the spectral index $n_B$.
Denoting the spectrum as function of scale
by $B(\lambda)=\sqrt{\left<B^2\right>_\lambda} \propto \lambda^{n_B}$,
the deflection angle by coherent components
can be expressed as $\theta_{\rm coh}\propto B(\lambda_{\rm IC})\lambda_{\rm IC}$,
while that by tangled components is
$\sqrt{\left<\theta_{\rm rand}\right>^2}\propto B(r_c)\sqrt{r_c\lambda_{\rm IC}}$
(for $r_c<\lambda_{\rm IC}$).
From these relations, $\theta_{\rm coh}/\theta_{\rm rand}=\left(\lambda_{\rm IC}/r_c\right)^{n_B+1/2}$,
so that tangled fields dominate when $n_B<-1/2$.
The time-dependent spectra discussed in \S \ref{section:results} for tangled and coherent fields
should correspond to cases with $n_B \ll -1/2$ and $n_B \gg -1/2$, respectively.

Let us now briefly discuss the observational prospects.
If magnetic fields in intergalactic regions of $\gamma$-$\gamma$ $e^\pm$ production are too strong,
the time delay can become so large that the flux of the delayed emission
becomes too low to be detectable by GeV-TeV instruments.
This implies an upper bound to the magnetic fields strengths that can be probed,
depending on the brightness of the GRB and the sensitivity of available observational facilities.
On the other hand, as mentioned in \S \ref{section:results},
a lower bound of $B \sim 10^{-20}$ G is set by the requirement that the time delay
is determined by magnetic deflection rather than by angular spreading from IC scattering.
Since the higher energy delayed photons generally come from higher energy $e^\pm$
that experience smaller deflections, they may be more suitable
for probing stronger magnetic fields in comparison to lower energy delayed photons.
Conversely, the lower energy delayed photons may have better chances to probe weaker fields.
Such issues related to observability will be addressed in more detail in a future publication \citep{2007obs}.

We emphasize that the geometry of the processes involved in delayed secondary emission
is naturally configured to probe only the magnetic fields
in intergalactic locations such as voids where they are likely to be weak,
without being affected by stronger fields existing around the structured regions of the universe.
First, the primary photons travel distances 
of $\lambda_{\rm IR}\sim 20$ Mpc $(n_{\rm IR}/0.1 {\rm cm}^3)^{-1}$,
the $\gamma$-$\gamma$ mean free path in the CIB,
depending moderately on their energy for $E_\gamma \sim 1$-10 TeV.
This should be far outside the GRB's host galaxy,
and presumably out into intergalactic void regions away from large-scale structure.
Furthemore, once they turn into $e^\pm$ and become subject to magnetic deflections,
they cool relatively quickly and stop radiating on scales of $\lambda_{\rm IC}\sim 35$ kpc $(E_{e}/10{\rm TeV})^{-1}$,
so that the associated secondary emission is sensitive only to the fields within the void regions.
Finally, the secondary photons will be indifferent to any foreground extragalactic or Galactic magnetic fields
as long as they do not induce further $e^\pm$ generation (see below).
These features are of great advantage over Faraday rotation methods,
which can only provide measures integrated along the line of sight
and are always contaminated by the Galactic contribution.

It remains to be seen, however, whether the field strengths in such intergalactic voids
are in the observationally favorable range of $B \sim 10^{-20}$-$10^{-17}$ G as discussed here.
It could plausibly be the case depending on their physical origin at high redshift
\citep[e.g.][]{2000ApJ...539..505G,2005A&A...443..367L,2007astro.ph..1329I,2007arXiv0710.4620T},
as long as they are not significantly polluted later by magnetization from
astrophysical sources such as galactic winds \citep{2006MNRAS.370..319B}
or radio galaxies \citep{2001ApJ...556..619F}.
We stress that such weak intergalactic fields would be very difficult to constrain observationally
other than through the delayed GeV-TeV emission.

\begin{figure}[ht]
\vspace*{0.8cm}
 \centering
  \rotatebox{0}{\includegraphics[width=0.7\textwidth]{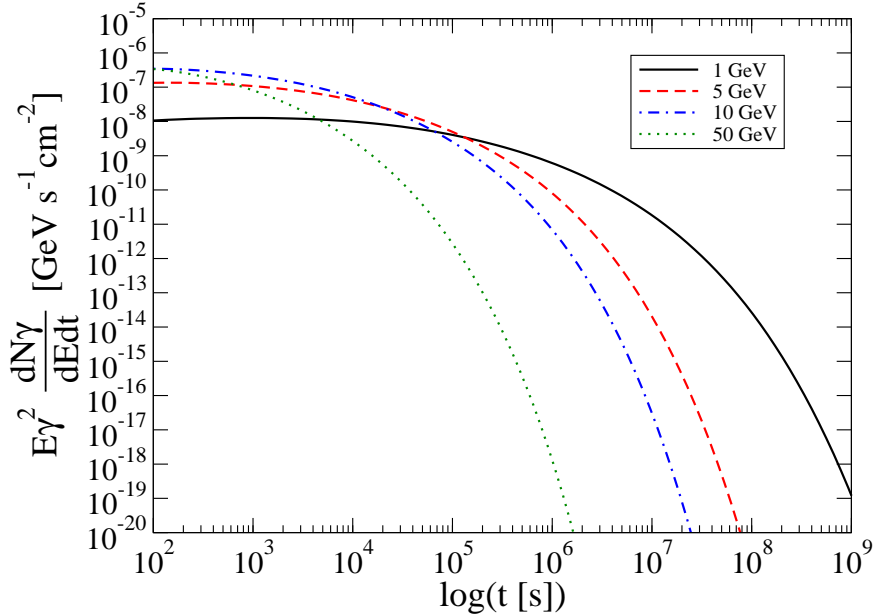}}
\caption{Light curves of the delayed emission at $E_\gamma =$ 1, 5, 10, and 50 GeV,
for a fiducial GRB at $z=1$, for tangled magnetic fields with $B=10^{-18}$ G and $r_c=100$ pc.}
\label{fig:lightcurve}
\end{figure}

In relation to the observability,
we caution that the GRB itself can possess long-lasting afterglow emission in the GeV-TeV bands
\citep[e.g.][]{1994MNRAS.269L..41M,1998ApJ...499L.131B,2001ApJ...559..110Z,2003ApJ...583..379I},
which may potentially mask the delayed secondary emission of our interest here.
However, the delayed emission should be distinguishable from the afterglow 
through clear differences in their time evolution.
Fig. \ref{fig:lightcurve} displays the light curves of the delayed emission for the case of Fig. \ref{fig:prompt},
which decay exponentially at late times corresponding to the typical delay time
$\Delta t_{\rm B}=(t_{\rm IC}+\lambda_{\rm IR}/c)\theta_B^2/2$,
with a characteristic energy dependence.
In contrast, power-law declines are generally expected for afterglow light curves until late times,
apart from possible breaks due to jet expansion, etc.
If the GeV-TeV afterglow is dominated by IC emission from the forward shock,
which could be the most common case \citep[e.g.][]{2001ApJ...559..110Z},
simultaneous measurements of the synchrotron emission in the radio to X-ray bands
will also provide tight constraints on the behavior of the high energy afterglow component.

\begin{figure}[ht]
\vspace*{0.8cm}
\centering
  \rotatebox{0}{\includegraphics[width=0.7\textwidth]{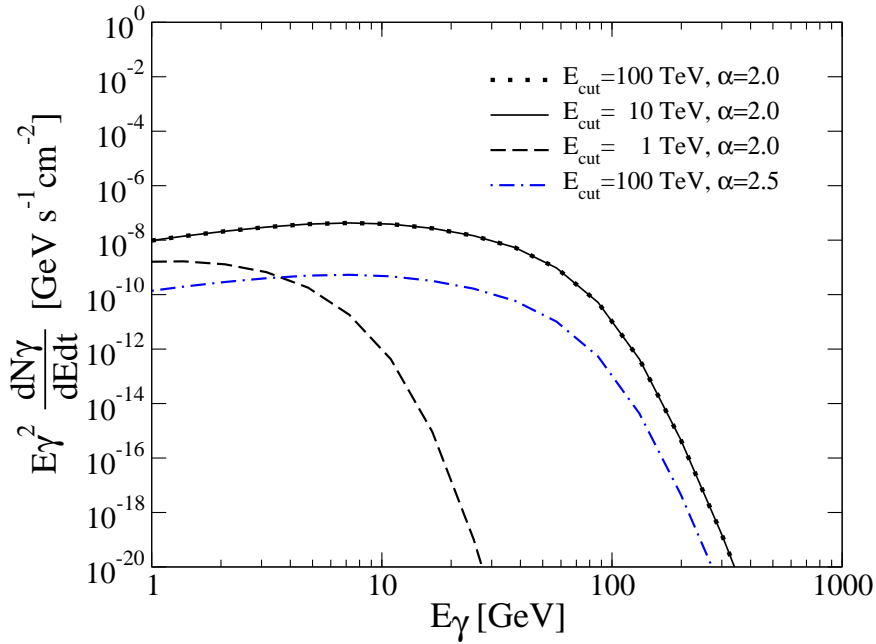}}
\caption{Spectra of delayed emission for a GRB at $z=1$ and $t_{\rm obs}=10^4$ s
in tangled magnetic fields with $B=10^{-18}$ G and $r_c=100$ pc
for different $\alpha$ and $E_{\rm cut}$ as indicated,
other GRB parameters remaining fiducial (Eq.(\ref{eq:GRBparameter})).}
\label{fig:varyGRB}
\end{figure}

A further caveat regards uncertainties in the spectrum of the primary emission.
Fig.(\ref{fig:varyGRB}) shows how the delayed emission spectra
depends on $\alpha$ and $E_{\rm cut}$ of the GRB prompt emission
as described by Eq.(\ref{eq:Band_Spectrum}).
Varying $\alpha$ and/or $E_{\rm cut} \la 10$ TeV have large effects,
as it changes the amount of primary photons above the threshold
for $\gamma$-$\gamma$ interactions with the CIB.
It appears to be insensitive to changes of $E_{\rm cut} \gtrsim 10$ TeV,
but this is actually an artifact of our approximation where we neglected
the effect of cascading beyond the second generation of $e^\pm$ production,
which could in reality be important for high $E_{\rm cut}$.
Such multi-generation cascades entails a complicated geometry
that cannot be treated analytically and requires numerical studies.
We remark, however, that $E_{\rm cut}$ is probably not much greater than $\sim 10$ TeV,
or else it would imply uncomfortably large bulk Lorentz factors for the GRB outflow,
independent of the emission mechanism
\citep[e.g.][]{2001ApJ...555..540L,2006ApJ...650.1004B}.
Note also that internal shock models generally predict separate, high-energy spectral components
above a simple power-law extrapolation from low energies \citep[e.g.][]{2004IJMPA..19.2385Z,2007arXiv0705.2910A}
which would enhance the delayed emission,
so the characterization of Eq.(\ref{eq:Band_Spectrum}) may be too simplistic.
More detailed and quantitative studies of the problem
including such realistic aspects will be presented elsewhere \citep{2007obs}.

Finally, we comment that the present approach to probing intergalactic magnetic fields
may also be applicable in some instances to flaring emission from blazars \citep{2002ApJ...580L...7D,2004A&A...415..483F}.
Recent observations of a nearby blazar by the H.E.S.S. collaboration
have revealed extraordinary flaring activity on minute timescales
with spectra extending to several TeV \citep{2007ApJ...664L..71A}.
Searching for delayed GeV-TeV emission from such objects may offer an important probe
of weak intergalactic magnetic fields in foreground voids.
Although their nontrivial light curves pose complications,
blazars have some crucial advantages over GRBs.
Besides the much larger photon statistics achievable,
the intrinsic primary emission before $\gamma$-$\gamma$ attenuation can be much better constrained
by virtue of simultaneous observations of the lower energy, synchrotron component \citep{1999ApJ...521L..33C}.

Observational tests of intergalactic magnetic fields through GeV-TeV gamma-ray astronomy should be forthcoming
with the anticipated launch of GLAST,
and ongoing developments with ground-based instruments such as
MAGIC, H.E.S.S., VERITAS, CANGAROO III, MILAGRO, etc.
The typical fluxes discussed above may be particularly interesting
for the upgraded MAGIC II and H.E.S.S. II telescopes with their lower threshold energies, 
not to mention future projects such as CTA and 5@5 (Fig. \ref{fig:varyz}).
In-depth discussions of the detectability with such facilities is deferred to future work.

\section{Summary}

With a new formulation that accounts for the proper geometry of the problem,
we have explored in some detail the delayed, secondary GeV-TeV emission from GRBs
and its potential to probe key properties of weak intergalactic magnetic fields.
The delayed emission was found to possess unique characteristics
that depend strongly on the amplitude as well as coherence length of the magnetic fields.
In the case of intergalactic fields with strength $B \sim 10^{-18}$ G and coherence length $r_c \sim 100$ pc,
the flux expected from a typical GRB at $z \sim 1$
is $\sim 10^{-8}$ GeV cm$^{-2}$ s$^{-1}$ at $\sim 10^4$ s after the burst,
in the range of interest for current and near future observational instruments.
GeV-TeV observations of GRBs may soon constrain the weakest intergalactic magnetic fields,
which could be difficult to do in any other way,
offering valuable insight into the physics of the intergalactic medium at high redshift,
and possibly into cosmological processes in the early universe.

\acknowledgements

We are very grateful to Tanja Kneiske for making her CIB models available on her website.
KT thanks K. Murase and S. Nagataki for helpful suggestions and useful discussions.
SI thanks P. Coppi for valuable comments.
KT and KI are supported by a Grant-in-Aid for the Japan Society for the Promotion of Science Fellowship.


\bibliography{grb}

%
%
\end{document}